\documentclass[useAMS,usenatbib]{mn2e}
\usepackage{epsfig}
\usepackage{threeparttable}
\usepackage{amsmath}

\title[Strange Dwarf Scenario for the Formation of SDSS J125733.63]
{A Strange Dwarf Scenario for the Formation of the Peculiar Double White Dwarf Binary SDSS J125733.63+542850.5}
\author[L. Jiang et al.]{Long Jiang$^{1,2}$, Wen-Cong Chen$^{1,2}$\thanks{E-mail:chenwc@pku.edu.cn}, Xiang-Dong Li$^{2,3}$ \thanks{E-mail:lixd@nju.edu.cn}\\
$^1$ School of Physics, Shangqiu Normal University, Shangqiu 476000, China;\\
$^2$ Key Laboratory of Modern Astronomy and Astrophysics (Nanjing University), Ministry of Education, Nanjing 210093, China;\\
$3$ Department of Astronomy, Nanjing University, Nanjing 210046, China}

\begin{document}

\date{}

\pagerange{\pageref{firstpage}--\pageref{lastpage}} \pubyear{2014}

\maketitle

\label{firstpage}

\begin{abstract}
The Hubble Space Telescope observation of the
double white dwarf (WD) binary SDSS J125733.63$+$542850.5
reveal that the massive WD has a surface gravity log$g_1\sim8.7$
(which implies a mass of $M_1\sim1.06~{\rm M_\odot}$)
and an effective temperature $T_1\sim13000$ K,
while the effective temperature of the low-mass WD ($M_2<0.24$ {\rm M$_\odot$}) is $T_2\sim6400K$.
Therefore, the massive and the low-mass WDs have a cooling age $\tau_1\sim1$ {\rm Gyr}
and $\tau_2\geq5$ {\rm Gyr}, respectively.
This is in contradiction with traditional binary evolution theory.
In this Letter, we propose a strange dwarf scenario to explain the formation of this double WD binary.
We assume that the massive WD is a strange dwarf originating from a phase transition in a $\sim1.1$ M$_\odot$ WD,
which has experienced accretion and spin-down processes.
Its high effective temperature could arise from the heating process during the phase transition.
Our simulations suggest that the progenitor of SDSS J125733.63$+$542850.5 can be
a binary system consisting of a $0.65~\rm M_{\odot}$ WD and a $1.5~\rm M_{\odot}$ main sequence star in a 1.492 day orbit.
Especially, the secondary star (i.e., the progenitor of the low mass WD) is likely to have an ultra-low metallicity of $Z=0.0001$.
\end{abstract}


\section{Introduction}
As the end products of binary evolution, double white dwarf (WD) binaries are
good probes for testing stellar and binary evolutionary theory \citep{mars95,toon14}.
Especially, they are thought to be
progenitors of Type Ia supernovae \citep{iben84, webb84},
AM CVn systems \citep{bree12, kili14}, and R CrB stars \citep{webb84}.
Furthermore, close double WDs are believed to be the main Galactic gravitational sources
in the frequency range of $10^{-4}$ to 0.1 Hz, which will be detected by the \emph{Laser Interferometer Space Antenna}
detector \citep{hils90,nele01, herm12}.

Based on the Sloan Digital Sky Survey (SDSS; \cite{eise06, york00}) subspectra,
the double WD binary SDSS J125733.63$+$542850.5 (hereafter J1257) was first
discovered by the Sloan WD Radial velocity Mining Survey \citep{bade09}.
Its radial velocity variations with a semi-amplitude of 323 $\rm km\,s^{-1}$ were interpreted to
originate from a 0.9 ${\rm M_\odot}$ WD while the companion was suggested to be a neutron star or black hole \citep{bade09}.
The two distinct components were revealed in the spectra
of \emph{B} and \emph{R} band spectroscopy,
and the  Balmer lines with deep radial velocity variable were identified to
come from a cool, extremely low-mass WD with mass less than 0.3 ${\rm M_\odot}$ \citep{kulk10, mars11}.

Recently, \cite{bour15} fit both the Hubble Space Telescope Cosmic Origins Spectrograph
and the Space Telescope Imaging Spectrograph spectra, and the SDSS \textit{ugriz} flux
with a Markov Chain Monte Carlo approach.
Their results indicate that the massive component has a surface gravity log $g_1\sim8.73\pm0.05$,
and an effective temperature $T_1\sim13030\pm70$ {\rm K}.
Detailed evolutionary models reveal the mass to be $M_1=1.06\pm0.05~{\rm M_\odot}$,
and a corresponding cooling age of $\tau_1=1.0$ {\rm Gyr} or 1.2 {\rm Gyr} for carbon/oxygen and oxygen/neon
WD models respectively \citep{kowa06,alth07,trem11}.
However, the low-mass WD with $M_2<0.24~{\rm M_\odot}$
has a temperature of $T_2\sim6400\rm ~K$ and a cooling age of $\tau_2\geq5$  Gyr \citep{mars11,bour15}.
The two cooling ages are in contradiction with binary stellar evolutionary theory.
The progenitor of the low-mass He WD probably had a mass of
$1-2~{\rm M_\odot}$ \citep{istr14}, and should have evolved much more slowly
than $5-6~{\rm M_\odot}$ progenitor of the more massive WD.
After the formation of the low-mass WD, CNO flashes could cause it to fill its Roche lobe,
and accretion heating would alter the thermal structure of the massive WD
with a duration of $\sim10^{6}~\rm yr$ \citep{bild06}.
However, this mass transfer timescale ($\sim100~\rm yr$) is too short to influence the cooling history of the massive WD.

In this paper, we propose that the difference in the cooling ages
originate from the heating process during the formation of a strange dwarf.
We describe the strange dwarf scenario in section 2.
In section 3, we discuss the changes of the orbital parameters during phase transition
and the possible spin-down process of the massive WD.
Employing the MESA code, we simulate the evolutionary history of J1257 in section 4.
A brief summary and discussion are presented in section 5.

\section{Strange dwarf Scenario}
Based on the hypothesis that the strange quark matter may be the most stable state of matter, \cite{witt84}
proposed that the pulsars may be strange quark stars (SSs) rather than neutron stars (NSs).
Following this idea, the concept of strange dwarfs (SDs) was introduced by \cite{glen95a, glen95b} as strange counterpart of WDs.
They pointed out that the inner density of ordinary stable WDs is always below critical density \textbf{$\rho_{\rm c}\sim10^9 ${\rm g~cm}$^3$},
which is the central density of a maximum-mass ($\sim1~{\rm M_\odot}$) WD \citep{baym71}.
The structure and thermal evolution of SDs with different masses have been studied by \cite{benv96} in detail.
Their computation indicates that the thermal evolution of SD with mass larger
than $1~{\rm M_\odot}$ is similar to that of a WD with the same mass.

Following these studies, we assume that if the central density of a WD exceeds a critical density, a strange quark core will emerge
in the center region and the star evolves to be a SD.
During this phase transition (PT, hereafter) process the mass of star will decrease slightly (about several percent).
Part of them will translate to the binding energy of the more compact SD,
and the other is lost from the star.
The energy transportation during the process may heat the star and result in a hotter SD.

Under the assumptions mentioned above,
we outline the evolutionary stages of J1257 as follows, which is also illustrated in Fig. 1.
\begin{itemize}
\item We start from a compact binary (with an orbital period $P_i\sim 1.5$ days),
which consists of a white dwarf with $M_1\sim0.6-0.8~{\rm M_\odot}$
and a main sequence star with $M_2\sim1.5~{\rm M_\odot}$.
\item Roche lobe overflow. The WD accretes around $0.3-0.5~{\rm M_\odot}$
material from the donor star, and spins up to nearly breakup rotation.
The spin-up process would also significantly broaden the hydrogen line profiles, as reported by \cite{kulk10}.
\item Formation of the second WD.
The mass transfer terminates and the secondary star evolves into a low-mass WD.
The central density of the primary WD,
which is centrifugally diluted by the fast spinning due to accretion starts to spin down.
Mean while the second WD gradually cools.
\item PT. After several ${\rm Gyr}$, the massive WD's spin down causes its density to be above the critical density,
thus PT takes place in its core and a nascent SD is heated up to $10^8~{\rm K}$.
\item Cooling of the massive SD.
According to \cite{benv96}, the cooling process of the SD is similar to a normal WD with the same mass.
\end{itemize}

\begin{figure}
\centering
\includegraphics[width=0.7\linewidth,trim={30 0 80 0}]{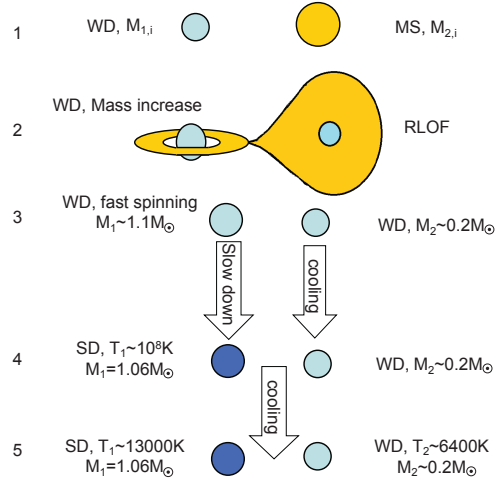}
\caption{\label{fig:sch} Illustration of our strange dwarf evolution from the compact white dwarf binary to the present observed system J1257.}
\end{figure}

The temperature of the nascent SD can be estimated as follows.
Assuming that the difference between the gravitational masses of a WD and a SD with the same quark number is $\Delta M$,
and that a fraction $\alpha$ ($<1$) of the rest energy of the mass loss during PT was used to heat the nascent SD,
the thermal energy received by the SD is:
\begin{equation}
Q=\alpha\Delta Mc^2.
\end{equation}

The internal energy of the SD with a temperature $T$ can be estimated as:
\begin{equation}
U_{\rm SD}=\frac{3}{2}{k_{\rm B}}NT,
\end{equation}
where $k_{\rm B}$ is Boltzmann's constant and $N=M_{\rm SD}/Am_{\rm u}$ is the total number of quarks and electrons.
Here $M_{\rm SD}$ is the mass of the SD, $m_{\rm u}$ is the atomic mass unit, $A$ is the relative particle mass.
Following \cite{alco86},  we assume that the SD is composed by roughly equal numbers of up, down and strange quarks and a small number (ignorable) electrons.
Considering that the mass of SD decreased a little during PT while the quark number keeps constant, we take $A\sim0.3$.

Since the effective temperature of the WD before PT was much lower than the nascent SD,
the internal energy of WD $U_{\rm WD}$ ($\ll U_{\rm SD}$) is ignorable.
Because $Q=U_{\rm SD} - U_{\rm WD}\approx U_{\rm SD}$, the temperature of the nascent SD can be written as:
\begin{equation}
T=\frac{2}{3}\frac{\alpha\Delta Mc^2}{k_{\rm B}N}=\frac{2}{3}\frac{\alpha\Delta Mc^2Am_{\rm u}}{k_{\rm B}M_{\rm SD}}.
\end{equation}

PT between NS and SS has been extensively investigated for different equations of state.
Several works suggested that the difference between the gravitational masses a NS and a
SS with the same baryon number is roughly $M_{\rm NS}-M_{\rm SS}\approx 0.15~{\rm M_\odot}$
for a NS with mass $\sim1.5~{\rm M_\odot}$ \citep[e.g.,][]{bomb00, drag07, marq17}\footnote{Lower value is also possible, for example,
the gravitational mass between NS and hyperon star given by \cite{scha02} is $\sim0.03~\rm {M_\odot}$.}.
Considering the difference in the mass and compactness between a WD and a NS, in this work we take
$\Delta M\sim 0.05~{\rm M_\odot}$.
Considering that most of the energy liberated during PT is assumed to be taken away by neutrinos (and anti-neutrinos),
similar as in supernova explosions \citep{kipp12}, and only a small fraction is used to heat the nascent SD, we set $\alpha\sim 0.001$ as lower limit.
Taking $A\sim0.3$, and $M_{\rm SD}=1.05~\rm M_{\odot}$, the nascent SD had an initial effective temperature $T\simeq10^8~{\rm K}$.
In comparison, the temperature of a $1.0 ~\rm M_\odot$ WD is $<10^7~\rm K$.

\section{Orbital Change during PT and the Spin Down of Progenitor WD}
\subsection{Orbital change during PT}

We first discuss the influence of PT on the eccentricity of the binary. Considering that the PT in the core of WD took place quickly,
and a kick velocity $V_{\rm k}$ was imparted to the new born SD, one can solve the orbital parameters during PT following \cite{shao16}.
Due to the mass transfer with a long duration, the orbit of the binary before PT can be thought to be circular.
Setting $\phi$ to be the positional angle of $V_{\rm k}$ with respect to the pre-PT orbital plane
and $\theta$  the angle between $V_{\rm k}$ and the pre-PT orbital velocity $V_{\rm 0}$ ($=(2\pi GM_{\rm 0}/P_{\rm orb, 0})^{1/3}$),
the ratio between the semi-major axes before and after PT is \citep{hill83, dewi03}:
\begin{equation}
\frac{a_0}{a}=2-\frac{M_{0}}{M_{0}-\Delta{M}}(1+\nu+2\nu~{\rm cos}~\theta),
\end{equation}
where $\nu=V_{\rm k}/V_{\rm 0}$, and
$M_{0}$ and $P_{\rm orb, 0}$ are the total mass of the binary and orbital period of the binary before PT, respectively.
Under the influence of mass loss and kick, the eccentricity after PT can be written as \citep{hill83, dewi03}:
\begin{multline}
1-e^2=\frac{a_{\rm 0}M_{\rm 0}}{a({M_{0}-\Delta{M})}}[1+ 2\nu~{\rm cos}~\theta \\+ \nu^2({\rm cos}^2\theta + {\rm sin}^2\theta~{\rm sin}^2\phi)].
\end{multline}

Taking $M_{0}=M_{2}+M_{1}=1.3~{\rm M_\odot}$, $\Delta{M}=0.05~{\rm M_\odot}$, $P_{\rm orb, 0}=0.22~{\rm day}$,
we simulated the possibility of small eccentricities ($e<0.01$) after PT for different $V_{\rm k}$ in the range of $0-50~{\rm km~s^{-1}}$.
For each $V_{\rm k}$, we set $10^7$ independent random values for ${\rm cos}\theta$ and $\phi$
of uniform distribution in the interval of -1 to 1 and 0 to $\pi$, respectively.
According to the observations of \cite{bade09} and  \cite{mars11},
the current orbit of J1257 is circular and a WD binary with $e<0.01$ could evolve into a circular orbit on a timescale of $\sim1\ {\rm G yr}$.
Fig.~2 shows the possibility distribution for the eccentricity less than 0.01 with different kick velocities.
When $V_{\rm k}\leq5~{\rm km~s^{-1}}$ and $V_{\rm k}\geq50~{\rm km~s^{-1}}$, the possibilities with $e<0.01$ are less than $0.1\%$ and $0.2\%$, respectively.
However, the relevant possibility is $\geq1\%$ when $6\leq V_{\rm k}\leq20~{\rm km~s^{-1}}$.
Especially, the relevant possibility is as high as $10\%$ for a kick velocity range of $8-9~\rm km\, s^{-1}$.
Similar to accretion induced collapse of neutron star \citep{hurl10}, the nascent SD should obtain a low kick velocity.
Therefore, PT process has a relatively large possibility to result in a nearly circular orbit.

According to the relation between the pre-PT and the post-PT orbital separation $a_0/(1+e)\leq a \leq a_0/(1-e)$ \citep{flan75},
one can derive the change of orbital period is  $\leq2\%$  when $e<0.01$.
Since the changes are relatively small, we ignore the orbital change of the binary during PT in our simulation.

\begin{figure}
\centering
\includegraphics[width=0.9\linewidth,trim={30 0 80 0},clip]{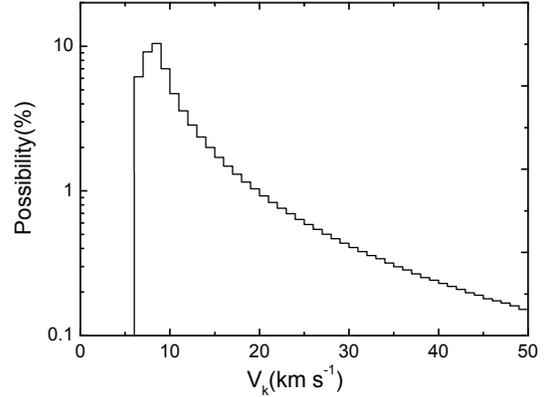}
\caption{\label{fig:sch} Possibility distribution for the eccentricity less than 0.01 under different values of the kick velocity.}
\end{figure}

\subsection{Spin down of massive WD }
Similar to pulsars, we consider the spin-down of WD with an angular velocity $\Omega=2\pi/P_{\rm s}$
is dominated by magnetic dipole radiation \footnote{Gravitational wave radiation might be an efficient mechanism
extracting angular momentum from fast rotating WDs due to r-mode instability
in a short timescale ($<10^{8}~\rm yr$, Yoon \& Langer 2004)}, and the energy loss rate is:
\begin{equation}
\dot{E}_{\rm d}=-\frac{2}{3c^{3}}\mu^{2}\Omega^{4},
\end{equation}
where $\mu=BR^3=B_{7}R_{9}^{3}\times 10^{34}~{\rm G~cm^3}$ is the magnetic dipole moment,
and $B_{7}$ is the surface magnetic field in units of $10^7~{\rm G}$, and $R_9$ is the radius in units of $10^9~{\rm cm}$ of the WD, respectively.

The rotational energy of WD changes at a rate:
\begin{equation}
\dot{E}_{\rm s}=I\Omega\dot{\Omega},
\end{equation}
where $I\sim MR^2\approx10^{51}R_{9}^{2}~{\rm g~cm}^{2}$ is the moment of inertia of WD.
If we assume that the braking torque of the WD fully originate from the magnetic dipole radiation, the spin period of the WD changes at a rate:
\begin{equation}
\dot{P}_{\rm s}=\frac{8\pi^2}{3c^3}\frac{\mu^2}{IP_{\rm s}}=K/{P_{\rm s}},
\end{equation}
where $K=8\pi^2\mu^2/3c^3I\sim B_{7}^{2}R_{9}^{4}\times{10^{-13}}{\rm ~s}$.

With simple integration, one can get the spin-down timescale of the WD from the initial spin period $P_{\rm s, 0}$
to the spin period of PT $P_{\rm s}$:
\begin{equation}
\tau_{\rm SD}=\frac{P_{\rm s}^{2}-P_{\rm s,0}^{2}}{2K}\approx \frac{P_{\rm s}^{2}-P_{\rm s,0}^{2}}{6B_{7}^{2}R_{9}^{4}}{\rm Myr}.
\end{equation}

Based on the theory of accretion disk-magnetic field interaction developed by \cite{GL1979},
\cite{kulk10} inferred that the magnetic field of J1527 was $\sim 10^{5}$ G when it spin up to its current spin period of $\sim60{\rm ~s}$.
However, \cite{cumm02} showed that rapid accretion could reduce the field strength at the surface of the accreting WD
because the field is advected into the interior by the accretion
flow. Therefore, many non-magnetic WDs ( $B\la
10^{5}$ G) may have submerged magnetic fields when they
were accreting at rates greater than the critical rate $\dot{M}_{\rm cr}=1-5\times 10^{-10}~\rm M_\odot\,yr^{-1}$,
and the magnetic field would re-emerge when the mass transfer terminates, and the re-emergence timescale of the field is:
\begin{equation}
\tau_{\rm re}\simeq 300\times(\frac{\Delta M_{\rm acc}}{0.1 {\rm ~M_\odot}})^{7/5}{\rm Myr},
\end{equation}
where $\Delta M_{\rm acc}$ is the accreted mass of the WD. Based on the model of \cite{cumm02} for magnetic field evolution, we propose that the surface field of the more massive WD in J1257 decreased from $\sim10^7 {\rm ~G}$ to $\sim10^5 {\rm ~G}$ due to rapid accretion.
According to the simulation in the next section, the accreted mass of the massive WD is $\Delta M_{\rm acc}\approx 0.45{\rm ~M_\odot}$, so the re-emergence timescale $\tau_{\rm re}\approx2.5\rm ~Gyr$.

Taking $P_{\rm s, 0}=10~{\rm s}$ (close to breakup rotation), $P_{\rm s}=60~{\rm s}$, $B_{7}=1$, and $R_{9}=0.6$,
one can derive the spin-down timescale of the massive WD is about $4.5~{\rm Gyr}$.
Considering the re-emergence timescale of the magnetic field $\tau_{\rm re}\approx 2.5~\rm Gyr$ ,
the total duration of 7 Gyr before the PT for the massive WD is approximately in agreement with the cooling age of the low mass WD.

\begin{figure}
\centering
\includegraphics[width=0.9\linewidth,trim={30 0 80 0},clip]{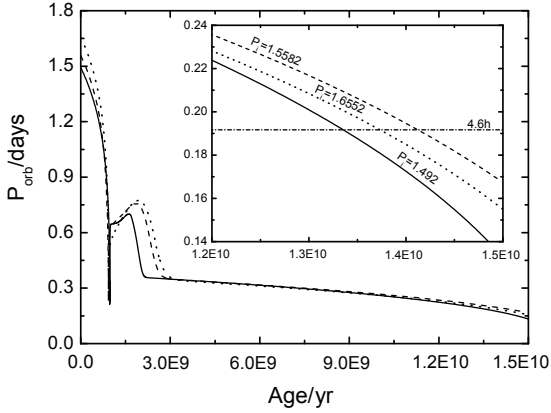}
\caption{\label{fig:sch} Evolution of orbital period for WD binaries including a $1.5{\rm~ M_\odot}$ donor star and a $0.65{\rm ~M_\odot}$ WD.
The solid, dashed and dotted curves represent the metallicity $z=0.0001,0.0005$ and $0.001$ respectively.
Numbers inside the curves denote the initial orbital periods.}
\end{figure}

\section{Numerical Simulation}
Using the MESA module \citep{paxt15}, we have simulated the evolution of the compact WD binary consisting of a WD and a main sequence star,
to test whether it is possible to reproduce the characteristics of J1257.
According to the estimation in the previous section, the orbital change during PT can be neglected while the mass growth of the WD is considered.

For an accreting WD, hydrogen and helium shell flashes always trigger nova outbursts,
which blow off the accreted matter and even result in convective dredge-up.
Therefore, the mass accumulation efficiency for accreting hydrogen should be less than 1.
During the mass transfer, the mass growth rate of the accreting WD is described as follows:
\begin{equation}
\dot{M}_{\rm 1}=\eta_{\rm He}\eta_{\rm H}|\dot{M}_{2}|,
\end{equation}
where $\dot{M}_{2}$ is the mass transfer rate of the donor star,
$\eta_{\rm H}$ and $\eta_{\rm He}$ is the accumulation efficiencies
during hydrogen burning and helium burning, respectively.

\begin{figure}
\centering
\includegraphics[width=0.9\linewidth,trim={30 0 80 0},clip]{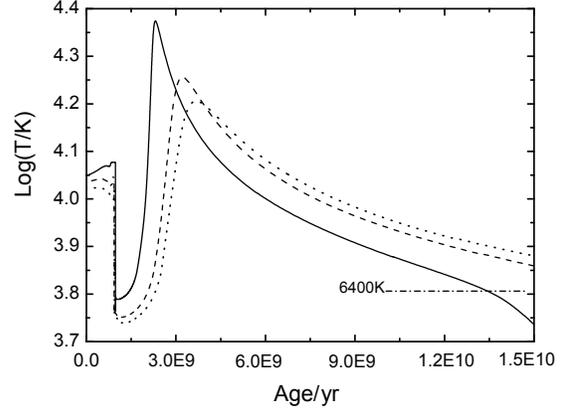}
\caption{\label{fig:sch} Evolutionary tracks of the effective temperature of donor star for WD binaries including a $1.5{\rm~ M_\odot}$ donor star
and a $0.65{\rm ~M_\odot}$ WD. Three cases are same to Figure 3.}
\end{figure}

For the accumulation efficiency of hydrogen, a prescription given by
\cite{hach99} and \cite{han04} was adopted, i.e.
\begin{equation}
\eta_{\rm H}=
\begin{cases}
\dot{M}_{\rm cr}/|\dot{M}_{2}|  &\text{$|\dot{M}_2|>\dot{M}_{\rm cr}$}, \\
1 &\text{$\dot{M}_{\rm cr}>|\dot{M}_2|>0.125\dot{M}_{\rm cr}$},\\
0 &\text{$|\dot{M}_2|<0.125\dot{M}_{\rm cr}$}.
\end{cases}
\end{equation}
In equation (12), $\dot{M}_{\rm cr}$ is a critical mass-accretion rate:
\begin{equation}
\dot{M}_{\rm cr}=5.3\times10^{-7}\frac{1.7-X}{X}(M_{1}-0.4)~{\rm M}_{\odot}~{\rm yr}^{-1},
\end{equation}
where $X$ is the mass abundance of hydrogen in the accreted matter.

For the accumulation efficiency $\eta_{\rm He}$ during helium burning, the prescriptions given by \cite{kato04} was adopted.
The mass loss $(1-\eta_{\rm H}\eta_{\rm He})\dot{M}_{2}$ during hydrogen and
helium burning is assumed to be ejected in the vicinity of the WD in the form of isotropic winds,
carrying away the specific angular momentum of the WD \citep{hach96, sobe97}.
In addition, we also consider angular momentum loss caused by gravitational radiation and
magnetic braking (with $\gamma=3.0$, \cite{rapp83, paxt15}).

\begin{figure}
\centering
\includegraphics[width=0.9\linewidth,trim={30 0 80 0},clip]{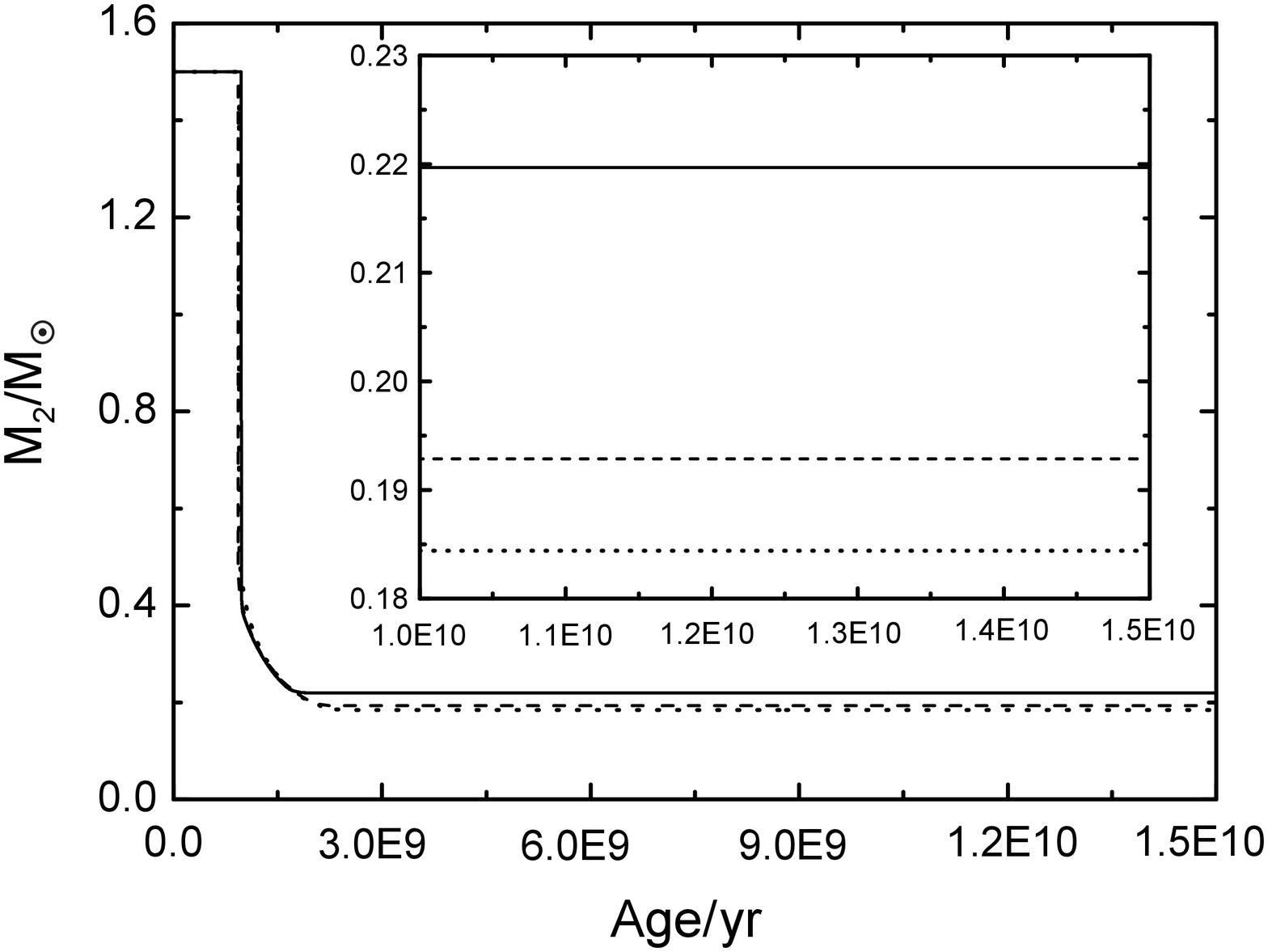}
\includegraphics[width=0.9\linewidth,trim={30 0 80 0},clip]{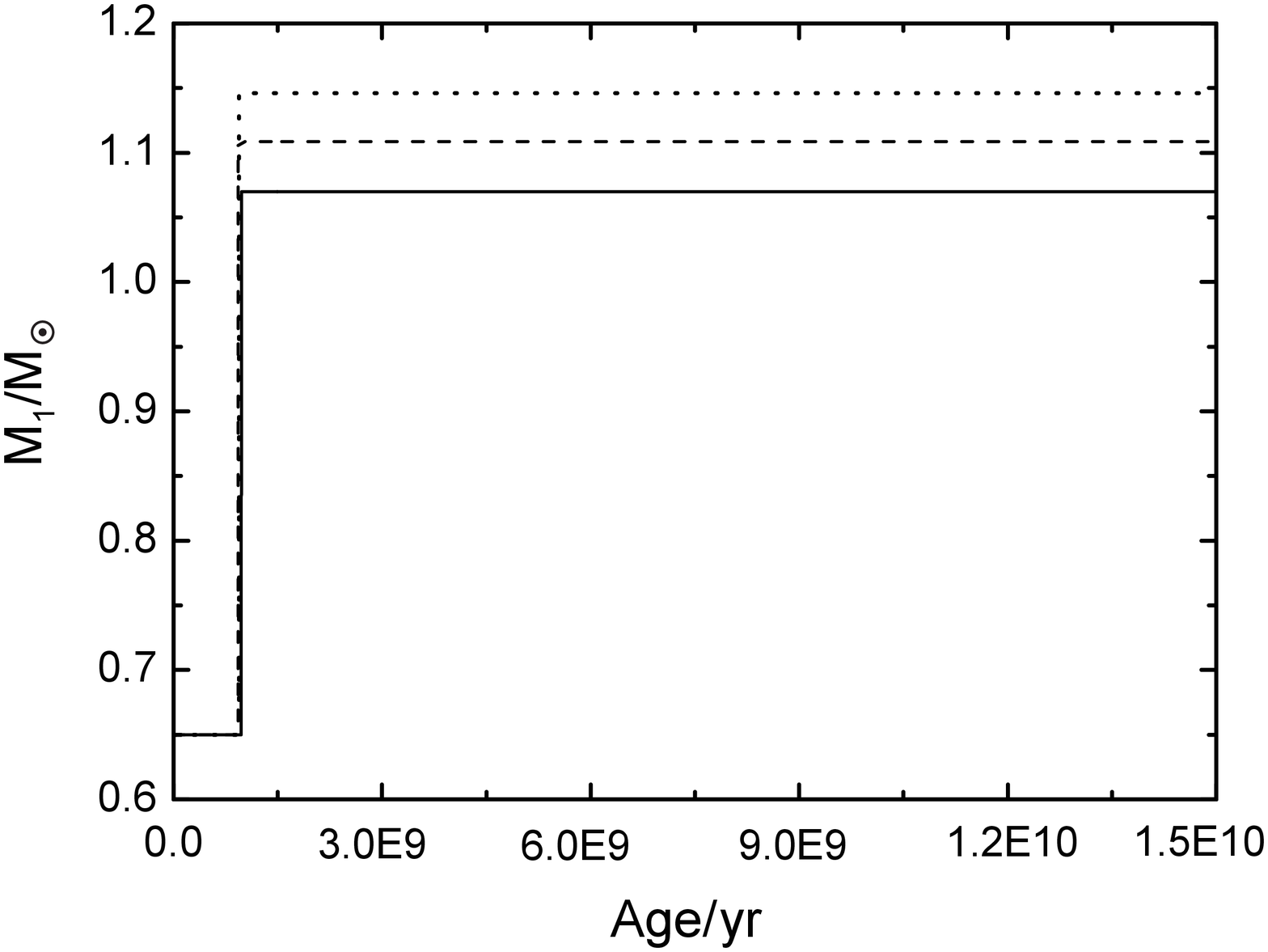}
\caption{\label{fig:sch} Evolution of the donor-star mass (top panel) and the WD mass (bottom panel) WD binaries including a $1.5{\rm~ M_\odot}$ donor star
and a $0.65{\rm ~M_\odot}$ WD. Three cases are same to Figure 3.}
\end{figure}

To study the progenitor properties of J1257, we simulated the evolution of a large number of WD binaries.
The relevant binaries would be thought to be the progenitor candidates of J1257 if the following three
conditions are satisfied: (1) the current orbital period is 4.6 hr when the age of the system is within Hubble time;
(2) the binary evolves into a detached system when the accreting WD's mass is about $1.05-1.15{\rm ~M_\odot}$
(Because the WD must experience a long-term spin-down process,
hence the donor star should not overflow its Roche lobe when the WD mass increases to $1.05-1.15~{\rm ~M_\odot}$);
(3) the effective temperature of the donor star is near 6400 K.
Figures 2-4 show an example evolution in which the initial donor mass  and the initial WD mass are $1.5{\rm~ M_\odot}$, and $0.65{\rm ~M_\odot}$, respectively.
We change the initial metallicity in order to fit the observed parameters of J1257.

As shown in Fig. 3, because the material is transferred from the more massive donor star to the less massive WD, the orbital period firstly decreases.
With the reversal of the mass ratio, the orbital period increases until the binary evolve into a detached system.
Subsequently, the donor star gradually evolve into a WD and enters the cooling stage,
and magnetic braking and gravitational radiation induce a compact double WDs binary.
Once the mass transfer ceases, the massive WD spin down due to magnetic dipole radiation, then trigger PT during the cooling of the low-mass WD.
The heating process during PT results in the formation of a hot SD.
Similar to Fig. 4, the simulated effective temperatures of the low-mass WD are always
higher than the observation in the Hubble time except for $z=0.0001$,
in which the donor star orbits a WD with an initial orbital period of 1.492 days.
Fig. 5 shows the evolutionary tracks of the donor-star mass and the accreting WD's mass.
It is clear that our simulated donor-star masses are consistent with the observed data for three different metallicities.
In calculation, the donor star with higher metallicity would produce systems with lower mass secondary WDs.
These difference should arise from the metallicity dependence of the stellar wind mass loss,
which tend to reduce the stellar mass of donor star.

\section{Summary and Discussion}
Assuming both WD and SD are different stages of stellar evolution,
in this work we propose a SD scenario to interpret the puzzle of the cooling age of two WD in J1257.
The massive WD is thought to be a SD originating from the PT of the $1.05-1.15~{\rm ~M_\odot}$ WD,
thus its higher effective temperature can be interpreted as a result of heating during PT.
A simple estimation indicates that a mass loss $\Delta M\sim 0.05~{\rm M_\odot}$ during PT can heated the nascent SD up to $10^8~{\rm K}$.
Based on these assumptions, we use the MESA code to simulate the evolution of a large number of WD binaries consisting of a $0.65~{\rm ~M_\odot}$ WD
and a $1.5~{\rm M_\odot}$ main-sequence star for different initial orbital period and metallicities.
Our simulation indicate that metallicities have important influence on the effective temperature of the donor star.
When $z=0.0001$, the calculate orbital period, the donor-star mass, and the effective temperature of the donor star are consistent with the observed data.
Therefore, we propose that the PT of a massive WD may be responsible for the puzzling cooling age of two WDs.
We expect further detailed multi-waveband observations for this source to obtain more precise constraints.

\section*{Acknowledgments}
This work was supported by the National Natural Science Foundation of China under grant number 11573016, 11733009, 11773015 and 11605110,
the National Key Research and Development Program of China (2016YFA0400803), and the Program for Innovative Research Team
 (in Science and Technology) at the University of Henan Province.

\bsp

\label{lastpage}

\end{document}